\documentclass[preprint,12pt]{elsarticle}
\usepackage{graphicx}
\usepackage{graphicx}
\usepackage{lipsum}
\usepackage{amssymb}
\usepackage{amsmath}
\usepackage{color}
\usepackage{multirow}
\usepackage{afterpage}
\usepackage{dsfont}
\usepackage{bm}%
\def\fnum@figure{\figurename\thefigure}
\renewcommand{\figurename}{Fig.}

\journal{Phys. Lett. A}
\begin{document}

\title
 {Fidelity based Measurement Induced Nonlocality over two-sided measurements}
\author{R. Muthuganesan, R. Sankaranarayanan}
\address{Department of Physics, National Institute of Technology\\ Tiruchirappalli-620015, Tamil Nadu, India.}

\begin{abstract}
In this paper, we introduce quantum fidelity based measurement induced nonlocality for bipartite state over two-sided von Neumann projective measurements. While all the properties of this quantity are reflected from that of one-sided measurement, the latter one is shown to set an upper bound for arbitrary bipartite state. As an illustration, we have studied the nonlocality of Bell diagonal state.
\end{abstract}

\maketitle
\section{Introduction}

Quantum correlation, a key in understanding bipartite composite quantum system, is a resource for powerful applications of quantum information and quantum computation. Though entanglement  between the constituents of composite system is a kind of nonlocal correlation,  it could not account all forms of  correlation contained in a quantum system. Recently, the correlation measures beyond entanglement  such as quantum discord\cite{Ollivier2001}, quantum deficit \cite{defict2016}, measurement induced disturbance \cite{Luo2008}, uncertainty induced locality (UIN) \cite{UIN2014} have been introduced. Of which, computation of discord involves complex optimization procedure \cite{Girolami2011} and is an NP problem \cite{Huang}. 

Recently, the correlation based on distance measures have been paid wide attention due to their computabilty and experimental realization \cite{{Dakic2010}}. One such quantity is measurement induced nonlocality (MIN), which captures the nonlocal effects of a quantum state due to locally invariant projective measurements \cite{Luo2011}. Further, MIN is more useful in quantum cryptography, dense coding and remote state control \cite{{Schrodinger1936},{Wiseman2007},{Peters2005},{Mattle1996},{Li2002},{Xiang2005}}. MIN also has experimental realization in terms of local Pauli operators \cite{Girolami}. However, this quantity suffers from the so called local ancilla problem - change may be effected through some trivial and uncorrelated action of the unmeasured party \cite{Piani2012}. This problem can be circumvented by replacing density matrix by its square root \cite{Chang2013}. In order to address this problem,  MIN has also been investigated in terms of relative entropy \cite{Xi2012}, von Neumann entropy \cite{Hu2012}, skew information \cite{Li2016}, Hellinger distance \cite{Chang2013} and trace distance \cite{Hu2015}. Recently, we have investigated the MIN based on quantum fidelity between pre-- and post--measurement state \cite{RMG1, RMG2}. We shall note that fidelity is also experimentally accessible using quantum networks \cite{Miszczak2009}, and hence nonlocal measure based on fidelity enjoys physical relevance.

In this article, we extend the fidelity based MIN for bipartite states over two-sided von Neumann projective measurements. It is shown that this quantity possesses all the properties as that of one-sided measurement,  including a remedy for local ancilla problem of MIN \cite{Chang2013}.  It is shown that the two-sided fidelity based MIN is bounded by one-sided counterpart for a $m \times n$ mixed bipartite state. The nonlocality of Bell diagonal state is also computed as an example.
\section{MIN over one-sided projective measurements}
Let us consider a bipartite quantum state $\rho $ in a Hilbert space $\mathcal{H}^a\otimes \mathcal{H}^b$. MIN is defined as the square of Hilbert--Schmidt norm of difference between pre-- and post--measurement state \cite{Luo2011} i.e., 
\begin{equation}
 N(\rho ) =~^{\text{max}}_{\Pi ^{a}}\| \rho - \Pi ^{a}(\rho )\| ^{2} 
\end{equation}
where the maximum is taken over the von Neumann projective measurements on subsystem $a$. Here $\Pi^{a}(\rho) = \sum _{k} (\Pi ^{a}_{k} \otimes   \mathds{1} ^{b}) \rho (\Pi ^{a}_{k} \otimes    \mathds{1}^{b} )$, with $\Pi ^{a}= \{\Pi ^{a}_{k}\}= \{|k\rangle \langle k|\}$ being the projective measurements on the subsystem $a$, which do not change the marginal state $\rho^{a}$ locally i.e., $\Pi ^{a}(\rho^{a})=\rho ^{a}$. If $\rho^{a}$ is non-degenerate, then the maximization is not required. In fact, the MIN has a closed formula for $2\times n$ dimensional states. The dual of this quantity is geometric measure of quantum discord (GD) of the given state $\rho$ which is defined as, 
\begin{equation}
 D(\rho ) =~^{\text{min}}_{\Omega \in \chi  }~\| \rho - \chi \| ^{2}  \nonumber
\end{equation}
where $\Omega$ is set of zero-discord state and $\chi=\sum_{i}p_{i}| \psi_{i}\rangle \langle \psi_{i}| \otimes \rho_{i}^b $ is a zero-discord state in the composite Hilbert space with probability distribution ${p_{i}}$. Due to the equivalence between the zero discord state $\chi $ and post- measurement state $\Pi ^{a}(\rho )$,  GD is reformulated as \cite{Luo2010pra}
\begin{equation}
 D(\rho ) =~^{\text{min}}_{\Pi ^{a}}\| \rho - \Pi ^{a}(\rho )\| ^{2}. \nonumber
\end{equation}

For nondegenerate $\rho_{a}$, both MIN and GD are equal. Both quantities are having wide range of applications due to their advantage of experimental realization. However, Hilbert-Schmidt norm based MIN could change due to  trivial and uncorrelated action on the unmeasured party $b$. This arises from appending an uncorrelated ancilla $c$ and regarding the state $\rho ^{a:bc}=\rho^{ab} \otimes  \rho ^{c}$ as a bipartite state with the partition $a$:$bc$; then it is easy to verify the following 
\begin{equation}
 N(\rho^{a:bc} ) = N(\rho^{ab})\text{Tr}(\rho ^{c})^2.  
\end{equation}
This relation implies that as long as $\rho^{c}$ is a mixed state,  MIN is altered by the addition of uncorrelated ancilla $c$ -- local ancilla problem. However, this problem can be remedied by replacing  $\rho$ by its square root. Nevertheless, different approaches on MIN have been adopted to resolve this issue as mentioned earlier. In what follows, we define MIN based on quantum fidelity between pre-- and post-- measured states, with measurement being performed on both the subsystems $a$ and $b$.

\section{MIN over two-sided projective measurements}
 One of the natural ways to define MIN based on fidelity induced metric (F--MIN) is \cite{RMG1, RMG2}
\begin{equation}
N^a_{\mathcal {F}}(\rho ) =~^{\text{max}}_{\Pi ^{a}}\mathcal {C}^{2}(\rho, \Pi ^{a}(\rho )) \label{min}  
\end{equation}
where  $\mathcal {C}(\rho ,\sigma )=\sqrt{1-\mathcal {F}(\rho ,\sigma )}$ is sine metric and 
\begin{equation}
\mathcal {F}(\rho ,\sigma )=\frac{(\text{Tr}(\rho \sigma ))^{2}}{\text{Tr}(\rho^{2})\text{Tr}(\sigma ^{2})}
\end{equation}
is the quantum fidelity between the states $\rho$ and $\sigma$ \cite{Wang2008}, which satisfies the axioms of original fidelity \cite{Jozsa1994}. In other words, MIN is defined in terms of the fidelity between pre-- and post-- measurement state. We have already shown that this quantity remedies the local ancilla problem of MIN. In the same spirit, to capture nonlocal effects of quantum state, we introduce fidelity based two-sided measurement induced nonlocality as 
\begin{equation}
N^{ab}_{\mathcal {F}}(\rho ) =~^{\text{max}}_{\Pi ^{ab}}\mathcal {C}^{2}(\rho, \Pi ^{ab}(\rho )) \label{2min}  
\end{equation}
where the maximum is taken over the von Neumann projective measurements on subsystems $a$ and $b$. Here $\Pi^{ab}(\rho) = \sum _{k,k'} (\Pi ^{a}_{k} \otimes   \Pi ^{b}_{k'}) \rho (\Pi ^{a}_{k} \otimes  \Pi ^{b}_{k'} )$, with $\Pi ^{a}= \{\Pi ^{a}_{k}\}= \{|k\rangle \langle k|\}$ and $\Pi ^{b}= \{\Pi ^{b}_{k'}\}= \{|k'\rangle \langle k'|\}$, which do not change the marginal states $\rho^{a}$ and $\rho^{b}$ respectively i.e., 
\begin{equation}
\Pi ^{a}(\rho^{a})=\rho^{a},   ~~~~~~~~~~  \Pi ^{b}(\rho^{b})=\rho^{b}.
\end{equation}

The two-sided MIN has the following properties:
\begin{enumerate}
\item[(i)]  $N^{ab}_{\mathcal {F}}(\rho )$ is non-negative i.e., $N^{ab}_{\mathcal {F}}(\rho )\geq 0$.
\item[(ii)] $N^{ab}_{\mathcal {F}}(\rho )=0$ for any product state $\rho=\rho_{a}\otimes  \rho _{b}$ and the classical state in the form $\rho =\sum _{i}p_{i}|i\rangle \langle i| \otimes   \rho_{i}  $ with nondegenerate marginal state $\rho^{a}=\sum_{i}p_{i}|i\rangle \langle i|   $. In both case, $\rho=\Pi ^{ab}(\rho )$, which leads to zero MIN. 
\item[(iii)] $N^{ab}_{\mathcal {F}}(\rho )$ is locally unitary  invariant for any unitary operators $U$ and $V$  in the sense that $N^{ab}_{\mathcal {F}}\left((U\otimes   V)\rho  (U\otimes   V)^\dagger\right)=N^{ab}_{\mathcal {F}} (\rho) $.
\item[(iv)] For any $2 \times n$ pure maximally entangled state $N^{ab}_{\mathcal {F}}(\rho) $ has the maximal value of $0.5$ (an immediate consequence of Theorem 1).
\item[(v)] $N^{ab}_{\mathcal {F}}(\rho )$ is invariant under the addition of any local ancilla to the unmeasured party (proof follows from multiplicative property of fidelity). 
\end{enumerate}

\section{MIN for pure state}
\label{sec:2}
\textit{{\bf Theorem 1:} For pure bipartite state with Schmidt decomposition $| \Psi \rangle =\sum_{i}\sqrt{s _{i}}| \alpha _{i} \rangle \otimes | \beta _{i}\rangle $  the two-sided F-MIN is }
\begin{equation}
N^{ab}_{\mathcal {F}}(| \Psi \rangle\langle \Psi| )=1- \sum_{i}s_{i}^{2}. \label{eq:pure}
\end{equation}

Proof is as follows: Here, von Neumann projective measurements expressed as $\Pi ^{ab}=\{\Pi ^{a}_{k}\otimes \Pi ^{b}_{k'}\} = \{| \alpha_{k}\rangle \langle \alpha_{k}|\otimes| \beta _{k'}\rangle \langle \beta _{k'}|\} $ do not alter the marginal states $\rho^{a}$ and $\rho^{b}$. Noting that
\begin{equation}
\rho=| \Psi \rangle \langle \Psi| = \sum_{ij}\sqrt{s_{i}s_{j}}| \alpha_{i} \rangle \langle \alpha_{j}| \otimes  | \beta_{i} \rangle \langle \beta_{j}| \nonumber  
\end{equation}
and the post-- measurement state becomes
\begin{equation}
\Pi^{ab}(\rho)= \sum_{k} s_{k}| \alpha_{k} \rangle \langle \alpha_{k}| \otimes  | \beta_{k} \rangle \langle \beta_{k}|. \nonumber  
\end{equation}
For any state $\rho$, we can easily show that $\text{Tr}(\rho~ \Pi ^{ab}(\rho))=\text{Tr}(\Pi ^{ab}(\rho))^2$. Since $\rho$ is pure, the fidelity between pre-- and post-- measurement state becomes $\mathcal {F}(\rho ,\Pi ^{ab}(\rho) )=\text{Tr}(\rho~ \Pi ^{ab}(\rho))$. Hence
\begin{equation}
N^{ab}_{\mathcal {F}}(| \Psi \rangle\langle \Psi| )=1- ~^\text{min}_{\Pi ^{ab}}~\text{Tr}(\rho~ \Pi ^{ab}(\rho))
\end{equation}
where the minimization is carried out over all possible projective measurements. After a straight forward simplification, we compute the fidelity between pre-- and post-- measurement state as
\begin{equation}
\mathcal {F}(\rho ,\Pi ^{ab}(\rho))=\sum_{k}s^{2}_{k}  \label{Fidelity}
\end{equation}
and hence the theorem is proved. 

Thus two-sided F--MIN for pure state coincides with that of one-sided F--MIN \cite{RMG1}, Hilbert-Schmidt norm \cite{Luo2011} and skew information \cite{Li2016} based MINs and remedied geometric discord \cite{Luo2010pra}. Eq. (\ref{Fidelity}) implies that the fidelity between pre-- and post-- measurement state is bounded by $1/m$ for any $m\times n$ dimensional state with $m\leq n$, and for pure maximally entangled state MIN reaches the maximal value $(m-1)/m$.  

Further, for any pure $2\times n$ dimensional system,  the trace distance based MIN $(N_{1})$ is given by
\begin{equation}
N_{1}(| \Psi \rangle\langle \Psi|)=2\sqrt{s_1s_2} \nonumber
\end{equation}
where $s_1+s_2=1$. For such a state the two sided F--MIN can be written as 
\begin{equation}
N^{ab}_{\mathcal {F}}(| \Psi \rangle\langle \Psi| )=(s_1+s_2)^2- (s^2_1+s^2_2) \nonumber 
\end{equation}
and hence,
\begin{equation}
N_{1}(| \Psi \rangle\langle \Psi|)=\sqrt{2N^{ab}_{\mathcal {F}}(| \Psi \rangle\langle \Psi| )}.
\end{equation}

\section{MIN for mixed state}

Let $\{X_{i}:i=0,1,2,\cdots,m^{2}-1\} \in \mathcal{B}(\mathcal{H}^a)$ be a set of orthonormal operators for the state space $\mathcal{H}^a$ with operator inner product $\langle X_{i}| X_{j}\rangle = \text{Tr}(X_{i}^{\dagger}X_{j})$. Similarly, one can define $\{Y_{j}:j=0,1,2,\cdots,n^{2}-1\}  \in \mathcal{B}(\mathcal{H}^b)$ for the state space $\mathcal{H}^b$. The operators $X_{i}$ and $Y_{j}$ are satisfying the conditions $\text{Tr}(X_{k}^{\dagger }X_{l})=\text{Tr}(Y_{k}^{\dagger}Y_{l})=\delta _{kl}$. With this one can construct a set of orthonormal operators $\{X_{i} \otimes Y_{j} \}\in \mathcal{B} (\mathcal{H}^{a}\otimes \mathcal{H}^{b}) $ for the composite system. Consequently, an arbitrary state of a bipartite composite system can be written as
\begin{equation}
\rho= \sum_{i,j}\gamma _{ij}X_{i}\otimes  Y_{j} \label{c}
\end{equation}
where $\Gamma = (\gamma _{ij} =\text{Tr} (\rho ~X_{i}\otimes  Y_{j}))$ is a $m^2 \times n^2$ real matrix.

After a straight forward calculation, the fidelity between pre-- and post-- measurement state is computed as
\begin{equation}
\mathcal{F}(\rho,\Pi^{ab}(\rho) )=\frac{\text{Tr}(\rho~\Pi^{ab}(\rho))}{\text{Tr}(\rho^2)}=\frac{\text{Tr}(A\Gamma B^t B\Gamma ^{t}A^{t})}{\| \Gamma \|^{2} } \nonumber
\end{equation}
where the matrix $A=(a_{ki}=\text{Tr}(|k \rangle \langle k| X_{i}))$  and $B=(b_{k'j}=\text{Tr}(|k' \rangle \langle k'| Y_{j}))$ are the rectangular matrices of order $m\times m^{2}$ and $n\times n^{2}$ respectively. Then, F-MIN is 
\begin{equation}
N^{ab}_{\mathcal{F}}(\rho)=\frac{1}{\|\Gamma  \|^{2} }\left[\|\Gamma  \|^{2}-^\text{min}_{A,B} \text{Tr}(A\Gamma B^t B\Gamma ^{t} A^{t})\right]. \label{result}
\end{equation}

If $X_{0}=\mathds{1}^{a}/\sqrt{m}$, $Y_{0}=\mathds{1}^{b}/\sqrt{n}$, and separating the terms in eq. (\ref{c}), the state $\rho$ can be written as  
\begin{equation}
\rho =\frac{1}{\sqrt{m n}}\frac{\mathds{1}^{a}}{\sqrt{m}}\otimes \frac{\mathds{1}^{b}}{\sqrt{n}}+\sum_{i=1}^{m^2 -1}x_{i}X_{i}\otimes\frac{\mathds{1}^{b}}{\sqrt{n}}+\frac{\mathds{1}^{a}}{\sqrt{m}}\otimes\sum_{j=1}^{n^2 -1}y_{j}Y_{j} +\sum_{i,j\neq 0}t_{ij }X_{i}\otimes Y_{j} \label{rho6}
\end{equation}
where   $x_{i}=\text{Tr}(\rho ~X_{i}\otimes \mathds{1}^{b})/\sqrt{n}$, $y_{j}=\text{Tr}(\rho ~\mathds{1}^{a} \otimes Y_{j} )/\sqrt{m}$ and $T = (t _{ij} =\text{Tr} (\rho ~X_{i}\otimes  Y_{j}))$ is a real correlation matrix of order $(m^{2}-1)\times(n^2 -1)$.\\

\textit{Theorem 2: F--MIN has a tight upper bound as }

\begin{equation}
N^{ab}_{\mathcal{F}}(\rho)\leq \text{max}~\{N^{a}_{\mathcal{F}}(\rho),N^{b}_{\mathcal{F}}(\rho)\} \label{FMIN}
\end{equation}
\textit{where  $N^{a(b)}_{\mathcal{F}}(\rho)$ is  MIN over one-sided projective measurements on subsystem $a(b)$.}

Proof is as follows: Following the optimization as given in \cite{RMG1}, we have an inequality for fidelity based MIN due to the projective measurements on $a$ as 
\begin{equation}
N^{a}_{\mathcal{F}}(\rho)\leq \frac{1}{\|\Gamma  \|^{2} }\left(\text {Tr}(\Gamma \Gamma ^t)-\sum_{i=1}^{m-1}\mu _{i}\right) \nonumber
\end{equation}
where $\mu _{i}$ are eigenvalues of matrix $\Gamma \Gamma ^t$,  listed in increasing order. In a similar way, for projective measurement on $b$, we have an upper bound as
\begin{equation}
N^{b}_{\mathcal{F}}(\rho)\leq \frac{1}{\|\Gamma  \|^{2} }\left(\text {Tr}(\Gamma \Gamma ^t)-\sum_{i=1}^{n-1}\mu _{i}\right). \nonumber
\end{equation}
Hence, from the above inequalities we have an upper bound for two-sided MIN from (\ref{result}) as
\begin{equation}
N^{ab}_{\mathcal{F}}(\rho)\leq \frac{1}{\|\Gamma  \|^{2} } \left(\text {Tr}(\Gamma \Gamma ^t)-\sum_{i=1}^ {\text{min}\{ { m-1,n-1 }\} }\mu _{i}\right) \label{FMIN1}
\end{equation}
and hence proved.

Alternatively, following the optimation procedure in \cite{RMG2}, the upper bounds for one-sided MINs are
\begin{eqnarray}
  N^{a}_{\mathcal{F}}(\rho)\leq \frac{1}{\|\Gamma  \|^{2} }\left(\text {Tr}S-\sum_{i=1}^{m-1}\lambda _{i}\right)   \label{upper1}\\ 
N^{b}_{\mathcal{F}}(\rho)\leq \frac{1}{\|\Gamma  \|^{2} }\left(\text {Tr}S-\sum_{i=1}^{n-1}\lambda _{i}\right).  \label{upper2}
\end{eqnarray}
Here matrix $S=\text{\bf x}\text{\bf x}^t+TT^t$ and $\lambda _{i}$ are the eigenvalues of $S$ listed in increasing order. Combining the Eqs. (\ref{upper1}) and (\ref{upper2}), we obtain another upper bound for two-sided F--MIN as 
\begin{equation}
N^{ab}_{\mathcal{F}}(\rho)\leq \frac{1}{\|\Gamma  \|^{2} } \left(\text {Tr}S-\sum_{i=1}^ {\text{min}\{ { m-1,n-1 }\} }\lambda_{i}\right) \label{FMIN2}
\end{equation}
and hence the theorem is proved. Due to Cauchy's interlacing theorem \citep{Bhatia1996}, we shall note that the above inequality is stronger than Eq. (\ref{FMIN1}). For $m=n$, $N^{ab}_{\mathcal{F}}(\rho)\leq N^{a}_{\mathcal{F}}(\rho) = N^{b}_{\mathcal{F}}(\rho)$.

For any $2 \times n$ dimensional state the fidelity based MIN $N^{a}_{\mathcal{F}}(\rho)$ is given as \cite{RMG2}
\begin{equation}
N^{a}_{\mathcal{F}}(\rho)= \frac{1}{\|\Gamma  \|^{2} }\left(\lambda_2 +\lambda_3 \right). \label{Final}
\end{equation}
It is the same for any $m \times 2$ state of one-sided fidelity based MIN $N^{b}_{\mathcal{F}}(\rho)$ over the projective measurement on $b$. Further, from Eq. (\ref{FMIN2})  it is easy to show that  for $2 \times 2$ dimensional state $N^{ab}_{\mathcal{F}}(\rho)=N^{a}_{\mathcal{F}}(\rho)=N^{b}_{\mathcal{F}}(\rho)$.

\textit{Theorem 3:  For $m\times n$ dimensional maximally entangled mixed state with $m\leq n$, we have }
\begin{align}
N^{a}_{\mathcal{F}}(\rho)=\frac{m-1}{m}, ~~~~~~~ N^{ab}_{\mathcal{F}}(\rho)=\frac{m-1}{m}. \nonumber
\end{align} 

Let $\rho$ be a maximally entangled mixed state expressed as 
\begin{align}
\rho=\sum_{k}p_k| \psi_{k} \rangle \langle \psi_{k} |, ~~~~~~~~\sum_{k}p_k=1, \nonumber
\end{align}
where $| \psi_{k}\rangle=\frac{1}{\sqrt{m}}\sum_{i}|i\rangle |i'\rangle $, ${|i\rangle }$ and ${|i'\rangle }$ are orthonormal bases of subsystems $a$ and $b$ respectively. The marginal state $\rho^a=\mathds{1}/m$ and hence any local projective measurements $\Pi ^{a}_{k} \otimes   \Pi ^{b}_{k'}$ and $\Pi ^{a}_{k}$ leaves $\rho^a$ invariant. From the  theorem 1, we have  
\begin{align}
\text{Tr}(|\psi_{k} \rangle \langle \psi_{k} |~\Pi^{ab}_\mathcal{F}(\psi_{k} \rangle \langle \psi_{k} |))=\sum_{i} s^2_i, \nonumber
\end{align}
and the fidelity between pre-- and post-- measurement state is 
\begin{align}
\mathcal{F}(\rho, \Pi^{ab}(\rho))=\frac{\text{Tr}\left(\sum_{k}p_k^2(|\psi_{k} \rangle \langle \psi_{k} |~\Pi^{ab}(\psi_{k} \rangle \langle \psi_{k} |))\right)}{\sum_{k}p_k^2}=\sum_{i} s^2_i, \nonumber
\end{align}
which is bounded by $1/m$, and hence the proof.
\section{Examples}
Here we study the F--MIN and MIN for well-known families of $2 \times 2$ mixed states namely, isotropic state, Werner state and Bell diagonal state. In the case of isotropic and Werner states, the results given for one sided F--MIN \cite{RMG2} hold good for two-sided F--MIN as well.

Next we consider the Bell diagonal state whose Bloch representation can be expressed as 
\begin{equation}
\rho^{BD}=\frac{1}{4}\left[\mathds{1}\otimes\mathds{1}+\sum^3_{i=1}c_i(\sigma_i \otimes \sigma_i)\right]
\end{equation}
where $\mathds{1}$ is identity matrix, $\sigma_i$ are Pauli spin matrices and  $-1\leq c_i=\langle \sigma_i \otimes\sigma_i \rangle \leq 1$. We shall note that the marginal state of Bell diagonal state is maximally mixed. In matrix form, 
\begin{equation}
\rho^{BD}=\frac{1}{4}
\begin{pmatrix}
1+c_3 & 0 & 0 & c_1-c_2 \\
0 & 1-c_3 & c_1+c_2 & 0  \\
0 & c_1+c_2 & 1-c_3 &0  \\
c_1-c_2 & 0 & 0 & 1+c_3
\end{pmatrix} . \nonumber
\end{equation}
As the name implies, Bell diagonal state has four maximally entangled Bell states as eigenvectors  corresponding to the eigenvalues 
\begin{equation}
\mu^{BD}_{i,j}=\frac{1}{4}\left[1+(-1)^i c_1-(-1)^{i+j}c_2+(-1)^j c_3\right]
\end{equation}
with $i,j=1,2$. If $\rho^{BD}$ describes a valid physical state, then $0\leq \mu^{BD}_{i,j}\leq 1$ and $\sum_{i,j}\mu^{BD}_{i,j}=1$. Under this constraint, the coordinates  $(c_1, c_2, c_3) $ must be restricted to the tetrahedron whose vertices situated on the points $(1, 1, -1)$, $(-1, -1, -1)$, $(1, -1, 1)$ and $(-1, 1, 1)$ representing Bell states (EPR pairs) \cite{Paula}. 

The MIN and F--MIN of Bell diagonal state are computed as 
\begin{align}
N(\rho^{BD})=\frac{1}{4}\left(\sum_{i=1}^{i=3}c_i^2-c_0^2\right),  ~~~~~~
N^{ab}_\mathcal{F}(\rho^{BD})=\frac{1}{\|\Gamma  \|^{2} } N(\rho^{BD})
\end{align}
where $c_0=\text{min}\{ \lvert c_1 \rvert, \lvert c_2 \rvert, \lvert c_3 \rvert \} $ and $\|\Gamma  \|^{2}=1+c_1^2+c_2^2+c_3^2$.  We shall note that MIN and F-MIN are maximum i.e., $ N(\rho^{BD})=N^{ab}_\mathcal{F}(\rho^{BD})=0.5$ in the above four vertices corresponding to Bell states.  On the other hand, both the MIN and F--MIN are vanishing for the coordinates $(0, 0, 0) $, at which the state $\rho^{BD}=\mathds{1}/4$ -- a maximally mixed state.
Further, for the symmetric states $c_1= c_2= c_3$  the  MINs are computed as 
\begin{equation}
 N(\rho^{BD})= \frac{1}{2}c_1^2  ~~~~~~~ N^{ab}_\mathcal{F}(\rho^{BD})=\frac{2c_1^2}{1+3c_1^2}
 \end{equation}
 with $0\leq c_1\leq 1/3$, and we observe that $N(\rho^{BD})\leq N^{ab}_\mathcal{F}(\rho^{BD})$.
\section{Conclusions}
In this article, we have proposed fidelity based measurement induced nonlocality (F--MIN) over two-sided measurements as a measure of quantum correlation for bipartite state. It is observed that this quantity possesses all the properties as that of F--MIN over one-sided measurement. In particular, the multiplicative property of fidelity is useful to resolve the local ancilla problem. It is shown that two-sided F--MIN coincides with one-sided F--MIN for arbitrary pure state. For $m \times n$ mixed state, the upper bound of two-sided F--MIN is obtained in terms of one-sided F--MIN. The MIN and F--MIN are also computed for the well-known Bell diagonal state.


\bibliographystyle{99}

\end{document}